%% file: main_llncs.tex
\documentclass[runningheads]{llncs}

\usepackage[T1]{fontenc}
\usepackage{graphicx}
\usepackage{amsmath,amssymb,amsfonts}
\usepackage{textcomp}
\usepackage{float}
\usepackage{tabularx}
\usepackage{bm}
\usepackage{enumitem}
\usepackage{algorithm}
\usepackage{algpseudocode}
\usepackage{listings}
\newfloat{listing}{htbp}{lol}
\usepackage{caption}
\usepackage[hidelinks,breaklinks]{hyperref}
\usepackage{subcaption}
\usepackage{listings}
\usepackage{seqsplit}
\usepackage{xurl}

\floatname{listing}{Code}
\newcommand{\ttbreak}[1]{\texttt{\seqsplit{#1}}}
\begin{document}

\title{MeReader: An Offline, Privacy-Preserving, Progress-Aware AI Assistant for Narrative eBook Reading\thanks{This preprint has not undergone peer review or any post-submission improvements or corrections. The Version of Record of this contribution is published in \emph{Advances in Computational Collective Intelligence} (ICCCI 2026), CCIS, vol.~3044, pp.~321--335, Springer, and is available online at \url{https://doi.org/10.1007/978-3-032-37936-8_22}.}}
\titlerunning{MeReader: Progress-Aware, Spoiler-Safe Reading Assistance}
\author{Mohammed Efaz\inst{1}\orcidID{0009-0002-3562-7607} \and
Udo Bub\inst{1}\orcidID{0000-0002-6018-2411} \and
Itilekha Podder\inst{1}\orcidID{0000-0001-7999-5597}}
\authorrunning{M. Efaz et al.}
\institute{Faculty of Informatics, Eötvös Loránd University (ELTE), Budapest, Hungary\\
\email{efaz@mohammedefaz.com, udobub@inf.elte.hu, itilekha19@inf.elte.hu}}
\maketitle

\begin{abstract}
This paper introduces MeReader, a privacy-preserving, offline and progress-aware AI assistant designed to function as a contextual companion for narrative reading on consumer hardware.
Unlike conventional Large Language Model (LLM)-based systems that process documents/books online in an all-texts-at-once type of way, thereby risking narrative spoilers, MeReader confines both retrieval and generation to the reader's distinct position in the text.
The system integrates a hybrid retrieval architecture, combining vector similarity, BM25, and summary embeddings, to ensure high fidelity to the reader's current context.
We evaluate the system across classical English books using both quantitative comprehension metrics and qualitative LLM-based independent judging.
The results identify a quality-latency frontier, showing that mid-sized local models (2b-4b) can achieve comprehension parity with larger counterparts (5b+) while maintaining acceptable response quality.
Importantly, the proposed progress-aware boundary mechanism is shown to prevent future content leakage without compromising retrieval quality and correctness.
These findings suggest that a localized, privacy-centric approach can augment, rather than replace, the digital reading experience, offering a good alternative to cloud-dependent services. 
Therein, there remains a lack of amalgamate systems which combine progress-aware, privacy-preserving, and narrative eBook reading with AI support.
It is in this very gap that the development of MeReader came to be; synthesizing said strands into an offline, book-constrained assistant focused on enhancing comprehension
without compromising narrative continuity or user privacy.
\end{abstract}

\section{Introduction}
\label{sec:introduction}
\input{chapters/01_intro}
\section{Related Work}
\label{sec:related_work}
\input{chapters/02_Related_Work}
\section{Proposed System Design}
\label{sec:system_design}
\input{chapters/03_System_Design}
\section{Experimental Setup}
\label{sec:implement}
\input{chapters/04_Implementation}
\section{Results and Discussion}
\label{sec:evaluation}
\input{chapters/05_Evaluation}

\section{Conclusion and Future Work}
\label{sec:conclusion}
\input{chapters/06_Conclusion}

\bibliographystyle{splncs04}
\bibliography{references}

\end{document}

%% file: chapters/01_intro.tex
Reading technologies have significantly reshaped the way individuals consume and interact with textual content. While conventional electronic book (eBook) readers have democratized access to literature, they remain largely passive platforms, offering limited support for deeper cognitive engagement beyond basic features such as highlighting, bookmarking, and keyword search~\cite{b1}. In parallel, the advent of powerful large language models (LLMs), such as OpenAI’s ChatGPT~\cite{b2} and Meta’s LLaMA series~\cite{b3}, has introduced highly interactive paradigms that allow users to upload complete documents and receive artificial intelligence (AI)-generated summaries or responses to natural language queries.

These systems offer efficiency, yet they displace the cognitive effort paramount for critical reading. By giving answers rather than facilitating discovery, they risk reducing readers to passive recipients. LLM-based tools that process entire texts at once often reveal spoilers, disrupting narrative continuity, while traditional eReaders lack support for tracking complex threads. This dual shortfall—AI that does too much and eReaders that do too little—showcases the need for a system that leverages AI to augment, rather than supplant, the reading experience.

In response, this study introduces MeReader, an offline, privacy-preserving, and progress-aware eBook reader designed to resolve the limitations of current market alternatives.
While "progress gating" exists in commercial ecosystems like Amazon Kindle's "Ask This Book"~\cite{engadget_kindle,verge_kindle}, it is a closed feature with proprietary formats and cloud services, both of which are antithetical to the philosophy and the aim of MeReader.
There exists also, in the realm of PDFs, tools like the Adobe Acrobat AI Assistant~\cite{verge_pdfs} that tackle primarily document summarization but their focus is more on productivity rather than narrative eBook reading; treating a novel like a database and querying answers and/or summaries based on prompts without regards to automatic spoiler prevention.
Open source alternatives, such as the ever popular KOReader~\cite{github_koreader} with its AI Assistant plugin~\cite{github_koreader_plugin}, prioritize privacy, but rely on a "trailing window" of recent text (limited to $~100$k characters) which in turn leads to an amnesia, of sorts, in its retrieval and generation system; the inability to recall plot points from early chapters once the reader advances deep in the book.
MeReader amalgamates these capabilities by generalizing the spoiler-safety of Kindle into an open architecture (one that is auditable) while also maintaining the deep, full-book retrieval memory lacking in the KOReader plugin.

This study aims to advance digital reading technologies by mitigating cognitive constraints associated with electronic text consumption, particularly reduced spatial memory and diminished recall, through the development of a minimally intrusive, context-sensitive AI assistant. The proposed system aims to maintain narrative coherence and reader agency while improving comprehension through localized, progress-aware support mechanisms. 
The remainder of this paper is organized as follows. Section~\ref{sec:related_work} reviews related work on AI-augmented reading tools, semantic retrieval architectures, privacy-preserving inference methods, and progress-aware user modeling. Section~\ref{sec:system_design} presents the design and architecture of MeReader, including EPUB parsing, text embedding, and system integration. Section~\ref{sec:implement} describes the technical implementation and software stack, detailing model optimization, UI/UX considerations, and inference logic. Section~\ref{sec:evaluation} reports the results of empirical evaluation across representative reader tasks. Section~\ref{sec:conclusion} discusses the broader implications of the findings and system limitations and concludes with a summary of contributions, design novelties, and avenues for future development.

%% file: chapters/02_Related_Work.tex
Recent advances in LLMs have demonstrated their potential as reading assistants. Chen and Leitch~\cite{b6} evaluated LLMs such as Claude for academic reading, reporting improvements in comprehension. However, they also highlighted limitations, including the models’ tendency to distract and concerns about privacy due to reliance on cloud infrastructure. Kryściński et al.~\cite{b7} introduced BOOKSUM, showing that LLMs can summarize long-form literary works such as novels and plays. Similarly, Spatharioti et al.~\cite{b8} found that LLM-enhanced web search improves the efficiency of information retrieval compared to traditional methods, particularly for consumer tasks. In the domain of document-level retrieval, Dense Passage Retrieval (DPR)~\cite{b9} was a significant step toward semantic search by learning dense representations of both queries and documents. However, as Reichman and Heck~\cite{b10} observed, DPR systems are typically trained on general corpora and are not easily adapted to specific, unseen documents. Improvements such as RocketQA~\cite{b11}, which leverages cross-batch negatives and distillation techniques, enhance retrieval quality but impose computational demands that limit applicability in low-resource or offline settings. Embedding models like nomic-embed-text-v2~\cite{b12}, built using a Mixture-of-Experts framework and trained on multilingual objectives, provide a balance between expressiveness and efficiency. Recent work, such as InfoCSE~\cite{b13}, has shown that masked contrastive objectives improve semantic alignment in retrieval tasks.

RAG has emerged as a robust strategy for combining semantic retrieval with generative inference. The canonical RAG architecture has been formalized and extended in works by Gao et al.~\cite{b14} and Wang et al.~\cite{b15}, including modular and task-adaptive variants. Vertical applications of RAG have been explored in domains such as medicine~\cite{b16,b17} and education~\cite{b18,b19}, typically in multi-document settings. However, user studies such as Levonian et al.~\cite{b19} indicate that retrieval strictness can conflict with user expectations, more so in educational contexts. Emerging architectures like FLARE~\cite{b20} introduce dynamic, content-aware retrieval strategies, though they often trade off control for flexibility. Privacy-preserving and local AI systems have gained attention due to increasing concerns over data security and user autonomy. Xu et al.~\cite{b21} demonstrated that on-device AI is feasible through techniques such as model quantization and runtime optimization. Lightweight infrastructure—including SQLite~\cite{b22}, Qdrant~\cite{b23}, and local inference frameworks like Ollama~\cite{b24}—supports practical deployment of vector search and inference on edge devices. However, as Mireshghallah et al.~\cite{b25} and Ippolito et al.~\cite{b26} showed, even pretrained models can leak memorized content, showing the importance of avoiding fine-tuning on user data. Similar privacy concerns were identified in on-device audio systems by Zhou et al.~\cite{b27}, who advocated for strictly offline processing to protect user interactions.

In summary, existing research has explored AI-assisted reading, semantic retrieval, RAG frameworks, and private inference independently. However, there remains a lack of integrated systems that combine progress-aware, privacy-preserving, and document-specific AI support within a single-reader context. This gap motivates the development of MeReader, which synthesizes these strands into an offline, book-constrained assistant focused on enhancing comprehension without compromising narrative continuity or user privacy.

%% file: chapters/03_System_Design.tex
MeReader follows a component-based architecture focusing on the separation of concerns and modularity. This approach allows individual components of the system to be developed, tested, and maintained independently, thereby reducing the risk of introducing regressions or conflicts in the core functionality during system evolution. It also facilitates easier scaling and integration of new features, ensuring that future enhancements can be incorporated with minimal disruption to the existing codebase. The design of MeReader integrates principles from both object-oriented programming (OOP) and service-oriented (SOA) architecture to establish robust and well-defined boundaries between system components.

\begin{figure}[!htbp]
  \centering
  \includegraphics[width=0.55\linewidth]{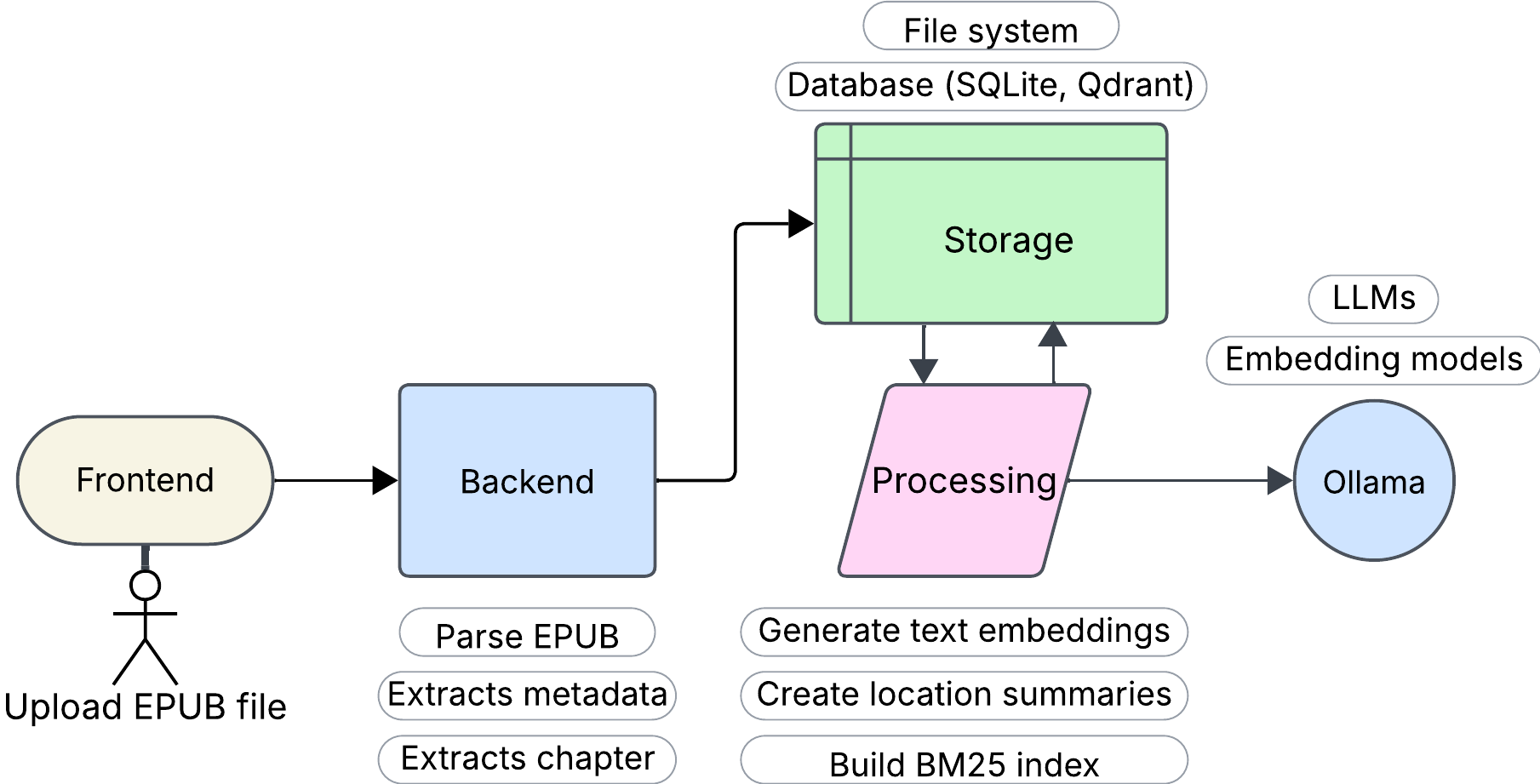}
  \caption{High Level Architecture}
  \label{fig:hla}
\end{figure}

The architecture, as shown in Figure~\ref{fig:hla}, is structured to promote clarity, with emphasis on maintaining distinct boundaries between components, thereby supporting modularity and ease of maintenance. The frontend of MeReader is developed using Tauri and Vue.js/JavaScript, leveraging a modular component structure to ensure a responsive and dynamic user experience. Each component is designed to handle a specific aspect of the user interface or interaction logic. The backend of MeReader is implemented using Python with FastAPI, providing a set of RESTful API endpoints that support the application’s core functionalities.
The backend is organized into specialized services, each responsible for distinct aspects of data processing, content management, and AI-driven features. 

MeReader utilizes a variety of storage solutions tailored to the specific needs of different data types within the system.
Each storage component is selected to optimize performance, reliability, and scalability for its designated role. Figure~\ref{fig:hla} also shows the journey of a book when it's first uploaded via the reader.
When a user uploads an EPUB file, (1) the original file is saved in a folder, (2) metadata is extracted, (3) content is processed and organized into chapters, (4) database records are created for the book and its structure, (5) text chunks are embedded and stored in the vector database and finally (6) content is made available for reading and querying.

\begin{figure}[!t]
  \centering
  \begin{minipage}[t]{0.48\columnwidth}
    \centering
    \includegraphics[width=\linewidth]{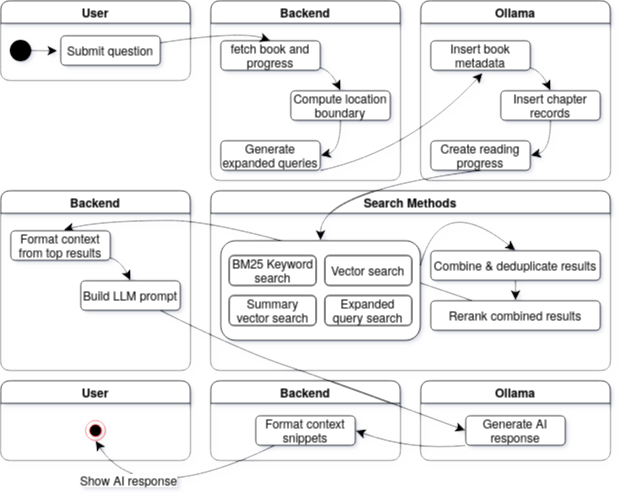}
    \caption{AI Query Flow}
    \label{fig:ai_flow}
  \end{minipage}\hfill
  \begin{minipage}[t]{0.48\columnwidth}
    \centering
    \includegraphics[width=\linewidth]{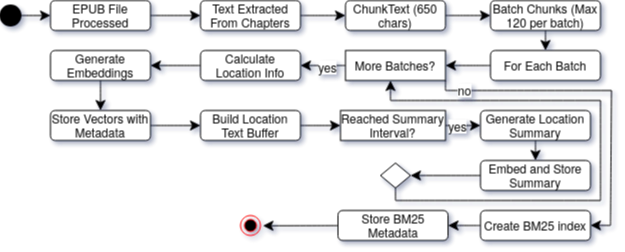}
    \caption{Embedding Pipeline}
    \label{fig:embedding_pipeline}
  \end{minipage}
\end{figure}

Figure~\ref{fig:ai_flow} depicts the end-to-end workflow for processing an AI query within MeReader. Upon query initiation, the system first retrieves the relevant book together with the user’s current reading position. To ensure contextual focus, a location boundary is then applied, restricting the search space to the most relevant sections of the text. Within this scope, multiple retrieval techniques are executed in parallel, including vector search, BM25 keyword search~\cite{b28}, and query expansion. The outputs of these methods are subsequently consolidated, with duplicates removed and results reranked. The highest-ranked passages are assembled into a contextual block, which is combined with the original query to construct the final prompt. This prompt is submitted to the LLM, whose response is returned to the user alongside the contextual information.

The embedding pipeline, as shown in Figure~\ref{fig:embedding_pipeline}, converts a given book text into vector representations for semantic search. Important steps include optimal chunking, metadata storing and batched processing, all of which elevate the quality of said embedding. Text flows through sequential processing stages: content extraction, chunking, batch organization, embedding generation, and vector storage. The pipeline also generates summary embeddings at regular intervals to enable higher-level content understanding and builds BM25 indices alongside vector embeddings to support keyword-based search capabilities.

%% file: chapters/04_Implementation.tex
We evaluated the MeReader architecture described in the previous section to assess its performance across content ingestion, query processing, and retrieval accuracy. The experiments were conducted on the deployed frontend–backend system, with evaluation focusing on component performance and end-to-end user experience.

The storage layer of MeReader employs a two-tier database strategy tailored to heterogeneous data access patterns; SQLite (via SQLAlchemy) for structured metadata and Qdrant for vector embeddings. This setup was benchmarked for scalability with respect to book size, embedding volume, and query throughput. The SQLite schema is normalized for reading applications and uses universally unique identifiers (UUIDs) as primary keys (PK). 
Foreign keys (FK) maintain referential integrity between entities.
\noindent Let $book \in \mathsf{Book}$ with attributes (id [PK], metadata, file\_ref, progress\_attr), $progress \in \mathsf{ReadingProgress}$ with attributes (id [PK], book\_id [FK], position, percentage, timestamp), and $chapter \in \mathsf{Chapter}$ with attributes (id [PK], book\_id [FK], order, content\_path, location\_bounds).

\noindent\textbf{Constraints:}
\begin{enumerate}[leftmargin=*,nosep]
    \item Each reading progress entry links to exactly one book (1:1):
    \[
        \forall progress,\ \exists! book:\ progress.book\_id = book.id
    \]
    \item Each chapter links to one book (1:*):
    \[
        \forall chapter,\ \exists book:\ chapter.book\_id = book.id
    \]
    \item On book deletion, associated reading progress entries and chapters are removed (cascade):
    \[
        \text{DELETE Book} \rightarrow \text{DELETE} \{\mathsf{ReadingProgress},\ \mathsf{Chapter}\}
    \]
\end{enumerate}

The core \textbf{Book} entity contains bibliographic metadata including publication details and cataloging identifiers, file system references for content and cover resources, and reading-specific attributes for progress calculation.
The \textbf{ReadingProgress} entity metadata maintains user state through current position tracking, completion percentages, and temporal access records.
The \textbf{Chapter} entity provides hierarchical content organization with sequential ordering and location boundaries defining chapter extents.

Qdrant provides specialized vector storage with cosine distance similarity for efficient semantic search operations.
Each book maintains a discrete collection enabling progress-aware filtering, with 768-dimensional vectors tagged by location identifiers for exact retrieval boundary enforcement.

\noindent Let $query\_vector$ be the embedding of the user query text, $filters$ the key--value constraints (e.g., book ID, location boundary), $candidates$ the set of retrieved segments matching the query and filters, $results$ the ordered list of candidates ranked by similarity score, and $top\_k(\cdot)$ the function returning the $k$ highest-scoring results.

\noindent\textbf{Vector Search Process:}
\begin{enumerate}[leftmargin=*,nosep]
    \item $query\_vector = embed\_text(user\_query)$
    \item $filters = \{book\_id: target\_book,\ location \le current\_position\}$
    \item $candidates = similarity\_search(
    query\_vector,\ filters,\allowbreak\ threshold)$
    \item $results = rank\_by\_cosine\_distance(candidates)$
    \item \textbf{return} $top\_k(results,\ limit)$
\end{enumerate}

The implementation achieves efficient similarity search while maintaining high recall rates for semantic queries through configurable score thresholds (0.6 for content retrieval, 0.65 for summary retrieval).

The storage architecture demonstrates efficient resource utilization.
The system processes content using configurable parameters of $400$-character chunks with $100$-character overlap, optimized for semantic coherence.
It comprises multiple components, including raw text content preservation for original document integrity, $768$-dimensional vector embeddings for semantic similarity search, BM25 inverted indices for keyword-based retrieval, and relational metadata for progress tracking and content organization. 
This hybrid approach achieves efficient resource usage, enabling scalable personal library management on consumer hardware while maintaining responsive query performance.

The content processing subsystem transforms uploaded EPUB files into searchable, semantically indexed representations while preserving reading progress boundaries essential for spoiler prevention (Algorithm~\ref{alg:semantic-boundary-detection}).
The implementation employs adaptive text segmentation algorithms that prioritize natural language boundaries over fixed size partitioning.

\noindent Let $F$ be the HTML content file, $chunk\_size = 400$ the target segment length, $overlap = 100$ the overlap between consecutive segments, and $min\_size = 100$ the minimum allowable segment length.

\begin{algorithm}[!t]
\caption{Semantic Boundary Detection}
\label{alg:semantic-boundary-detection}
\begin{algorithmic}[1]
\Require HTML content file $F$, $chunk\_size$, $overlap$, $min\_size$
\Ensure Semantically segmented text chunks $C$
\State $\mathit{text} \leftarrow \Call{ExtractTextFromHTML}{F}$
\State $\mathit{position} \leftarrow 0$;\quad $\mathit{chunks} \leftarrow [\,]$
\While{$\mathit{position} < \Call{Length}{\mathit{text}}$}
    \State $\mathit{buffer} \leftarrow \mathit{text}[\mathit{position} : \mathit{position} + 3 \times chunk\_size]$
    \State $\mathit{chunk\_end} \leftarrow \min(chunk\_size,\ \Call{Length}{\mathit{buffer}})$
    \Statex \Comment{Priority 1: paragraph boundaries}
    \State $\mathit{paragraph\_break} \leftarrow \Call{RFind}{\mathit{buffer},\ ``\backslash n\backslash n'',\ chunk\_size/2,\ chunk\_size+50}$    
    \If{$\mathit{paragraph\_break} \neq -1 \ \land\ \mathit{paragraph\_break} > chunk\_size/2$}
        \State $\mathit{chunk\_end} \leftarrow \mathit{paragraph\_break} + 2$
    \Else
        \Comment Priority 2: sentence boundaries
        \State $\mathit{sentence\_break} \leftarrow \Call{RFind}{\mathit{buffer},\ ``.\ '',\ chunk\_size/2,\ chunk\_size+30}$
        \If{$\mathit{sentence\_break} \neq -1 \ \land\ \mathit{sentence\_break} > chunk\_size/2$}
            \State $\mathit{chunk\_end} \leftarrow \mathit{sentence\_break} + 2$
        \EndIf
    \EndIf
    \State $\mathit{chunk} \leftarrow \Call{Trim}{\,\mathit{buffer}[0 : \mathit{chunk\_end}]\,}$
    \If{$\Call{Length}{\mathit{chunk}} \ge min\_size$}
        \State $\Call{Append}{\mathit{chunks},\ \mathit{chunk}}$
    \EndIf
    \State $\mathit{advance} \leftarrow \max\bigl(\mathit{chunk\_end} - overlap/2,\ chunk\_size/2\bigr)$
    \State $\mathit{position} \leftarrow \mathit{position} + \mathit{advance}$
    \If{$\mathit{position} \bmod (20 \times chunk\_size) = 0$}
        \State \Call{GarbageCollect}{}
    \EndIf
\EndWhile
\State \Return $\mathit{chunks}$
\end{algorithmic}
\end{algorithm}

The system implements location tracking using character-based calculations. Each location unit represents $1000$ characters, configurable via settings and enabling progress boundaries.

\begin{algorithm}[!t]
\caption{Progress-Aware Boundary Enforcement}
\label{alg:progress-aware-boundary}
\begin{algorithmic}[1]
\Function{CalculateSafeBoundary}{$current\_pos,\ total\_pos$}
    \If{$current\_pos \le 0$}
        \State \Return $1$
    \ElsIf{$current\_pos > total\_pos$}
        \State \Return $total\_pos$
    \Else
        \State \Return $current\_pos$
    \EndIf
\EndFunction
\end{algorithmic}
\end{algorithm}

As shown in Algorithm~\ref{alg:progress-aware-boundary}, the system calculates safe reading boundaries to prevent spoilers by ensuring users only access content they have already read.  
The boundary calculation function handles edge cases where users might be at the beginning or end of a book, ensuring valid location ranges for content filtering.

\noindent Completion percentage is computed as:
\[
    progress\_ratio \gets \min\!\bigl(100,\ \max(0,\ \frac{current\_position}{total\_positions} \times 100)\bigr)
\]

\begin{sloppypar}
Spoiler prevention is implemented via Qdrant filtering, e.g.
{\urlstyle{tt}\nolinkurl{filter_conditions.append(FieldCondition(key="location", range=Range(lte=boundary)))}},
which restricts results to content within the user’s current reading progress.
\end{sloppypar}

The system appends location based filters during vector searches, restricting results to current progress.
The embedding pipeline processes maximum $120$ chunks per batch with garbage collection every 20 chunks, generating summaries at $11$ location intervals.
The RAG architecture employs searches combining vector similarity, keyword matching, query expansion, and progress boundaries, using sequential execution with score normalization and fusion (Algorithm~\ref{alg:multi-modal-retrieval}).
Alternative queries from local LLMs complement retrieval (lines $20$–$26$), with the \texttt{DeduplicateAndRank} function removing semantically redundant results and orders the remaining items by their aggregated relevance score (line $28$).
Results exceeding eight items where queries contain more than three terms undergo LLM-based reranking on a ten point scale.

\begin{algorithm}[!t]
\caption{Multi-Modal Retrieval with Progress Filtering}
\label{alg:multi-modal-retrieval}
\begin{algorithmic}[1]
\Require Query $q$, book ID $B$, current location $L$, total locations $T$
\Ensure Ranked context passages $R$
\State $location\_boundary \gets \Call{CalculateBoundary}{L,\ T}$
\State $query\_embedding \gets \Call{EmbedText}{q}$
\State $expanded\_queries \gets$
\Statex \hspace{\algorithmicindent}\Call{GenerateAlternatives}{q,\ book\_title}
\State $results \gets [\,]$
\Statex
\Comment{Vector search (limit = 15, threshold = 0.6)}
\State $vector\_results \gets \Call{QdrantSearch}{query\_embedding,\ B,\allowbreak 15,\ 0.6,\ location\_boundary}$
\ForAll{$result \in vector\_results$}
    \State $result.method \gets$ ``vector''
\EndFor
\State $\Call{Extend}{results,\ vector\_results}$
\Statex
\Comment{BM25 keyword search (limit = 10)}
\State $bm25\_results \gets$
\Statex \hspace{\algorithmicindent}\Call{BM25Search}{q,\ B,\ 10,\ location\_boundary}
\ForAll{$result \in bm25\_results$}
    \State $result.method \gets$ ``bm25''
\EndFor
\State $\Call{Extend}{results,\ bm25\_results}$
\Statex
\Comment{Summary search (limit = 5, threshold = 0.65)}
\State $summary\_results \gets$
\Statex \hspace{\algorithmicindent}\Call{QdrantSearch}{query\_embedding,\ B,\ 5,\ 0.65,\ location\_boundary,\ ``summary''}
\ForAll{$result \in summary\_results$}
    \State $result.method \gets$ ``summary''
\EndFor
\State $\Call{Extend}{results,\ summary\_results}$
\Statex
\Comment{Expanded query search}
\ForAll{$expanded\_q \in expanded\_queries$}
    \State $expanded\_embedding \gets \Call{EmbedText}{expanded\_q}$
    \State $expanded\_results \gets$
    \Statex \hspace{\algorithmicindent}\Call{QdrantSearch}{expanded\_embedding,\ B,\ 5,\ 0.5,\allowbreak location\_boundary}
    \ForAll{$result \in expanded\_results$}
        \State $result.method \gets$ ``expanded\_vector''
    \EndFor
    \State $\Call{Extend}{results,\ expanded\_results}$
\EndFor
\State \Return $\Call{DeduplicateAndRank}{results}$
\end{algorithmic}
\end{algorithm}

The system applies method specific weights \(w_{\text{bm25}} = 0.40\),
\(w_{\text{vector}} = 0.60\),
\(w_{\text{summary}} = 0.40\),
and \(w_{\text{expanded}} = 0.70\).
These weights are applied after per-method score normalization to balance the contribution of keyword matching, semantic similarity, summary retrieval, and query expansion under a single ranking signal.
Each method’s raw scores are normalized by its maximum observed score for the current query, and the resulting normalized scores are then scaled by the corresponding \(w_{\text{method}}\) before being merged into a combined pool.

The final ranking is produced by sorting the merged candidates by their weighted normalized score, followed by the deduplication pass.
In this last step, \ttbreak{DeduplicateByContentSimilarity} removes semantically redundant passages whose overlap exceeds a predefined threshold, ensuring that only distinct context segments are retained while preserving the highest-scoring representative.

For model integration, MeReader leverages local language models via Ollama to enable privacy-preserving and offline operation.
Communication occurs through asynchronous HTTP API requests (default endpoint: \texttt{localhost:11434}), with configurable support for different model architectures.
The completion API exposes standard parameters such as temperature (default $0.7$), maximum tokens, and streaming options, while the embedding API uses \texttt{nomic-embed-text} to generate $768$-dimensional vectors with batched execution.

%% file: chapters/05_Evaluation.tex
\noindent
\begin{minipage}[t]{0.48\columnwidth}
\captionof{listing}{Comprehension example}
\label{code:comprehension-example}
\begin{lstlisting}[basicstyle=\ttfamily\scriptsize]
{
  "book": "1984",
  "query": "What are the three slogans of the Party?",
  "ground_truth": "WAR IS PEACE, FREEDOM IS SLAVERY, IGNORANCE IS STRENGTH"
}
\end{lstlisting}
\end{minipage}\hfill
\begin{minipage}[t]{0.48\columnwidth}
\captionof{listing}{Spoiler-prevention example}
\label{code:spoiler-example}
\begin{lstlisting}[basicstyle=\ttfamily\scriptsize]
{
  "book": "1984",
  "query": "What happens to the glass paperweight Winston bought?",
  "percentage_of_book_read": 55,
  "evaluation_rules": {
    "forbidden_keywords": ["smashed","pieces","hearth-stone","fragment of coral"]
  }
}
\end{lstlisting}
\end{minipage}

We evaluated a total of $600$ queries across three literary novels, (i) 1984 (George Orwell)~\cite{b31}, (ii) Of Mice and Men (John Steinbeck)~\cite{b32}, and (iii) The Death of Ivan Ilych (Leo Tolstoy)~\cite{b33}, using both technical and LLM-judged metrics.
The benchmark included 300 comprehension queries (30 queries per model across 10 models) in the format as shown in Code~\ref{code:comprehension-example}, and 300 spoiler prevention scenarios (30 queries per model across 10 models), in the format shown in  Code~\ref{code:spoiler-example}.
Each comprehension item paired a short reference answer generated with \texttt{Gemini 2.5 Pro}, while spoiler tasks defined “forbidden facts” with progress-based thresholds.
For every item, passages were retrieved through hybrid lexical–semantic search, deduplicated, and limited to six excerpts.
The models were instructed to answer only from the excerpts and to respond “Info not available in text so far” when necessary.
Responses were generated with temperature $0.3$.

\begin{figure}[!t]
  \centering
  \begin{minipage}[t]{0.48\columnwidth}
    \centering
    \includegraphics[width=\linewidth]{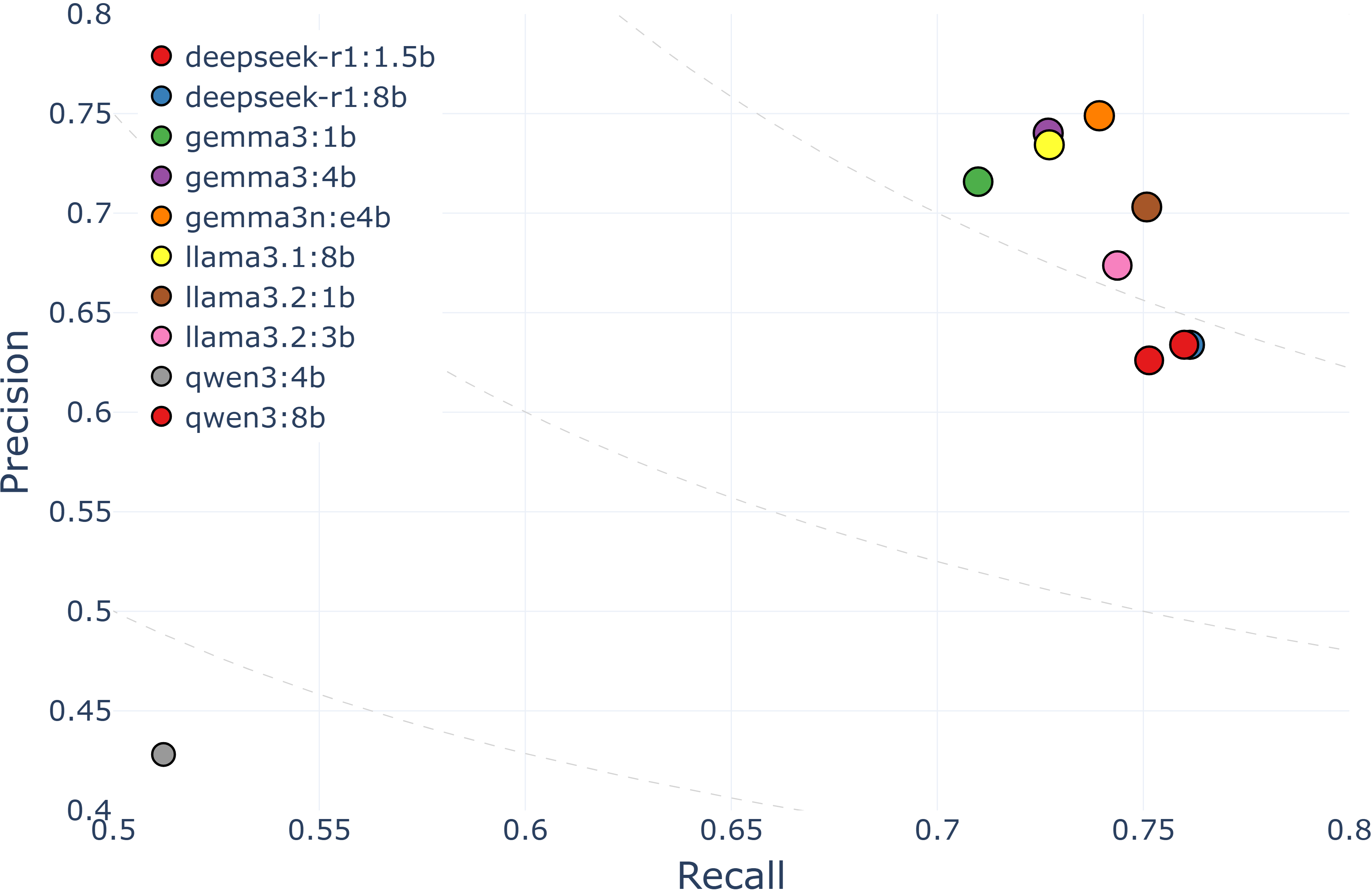}
    \caption{Precision Recall Plane}
    \label{fig:eval_precision_recall}
  \end{minipage}\hfill
  \begin{minipage}[t]{0.48\columnwidth}
    \centering
    \includegraphics[width=\linewidth]{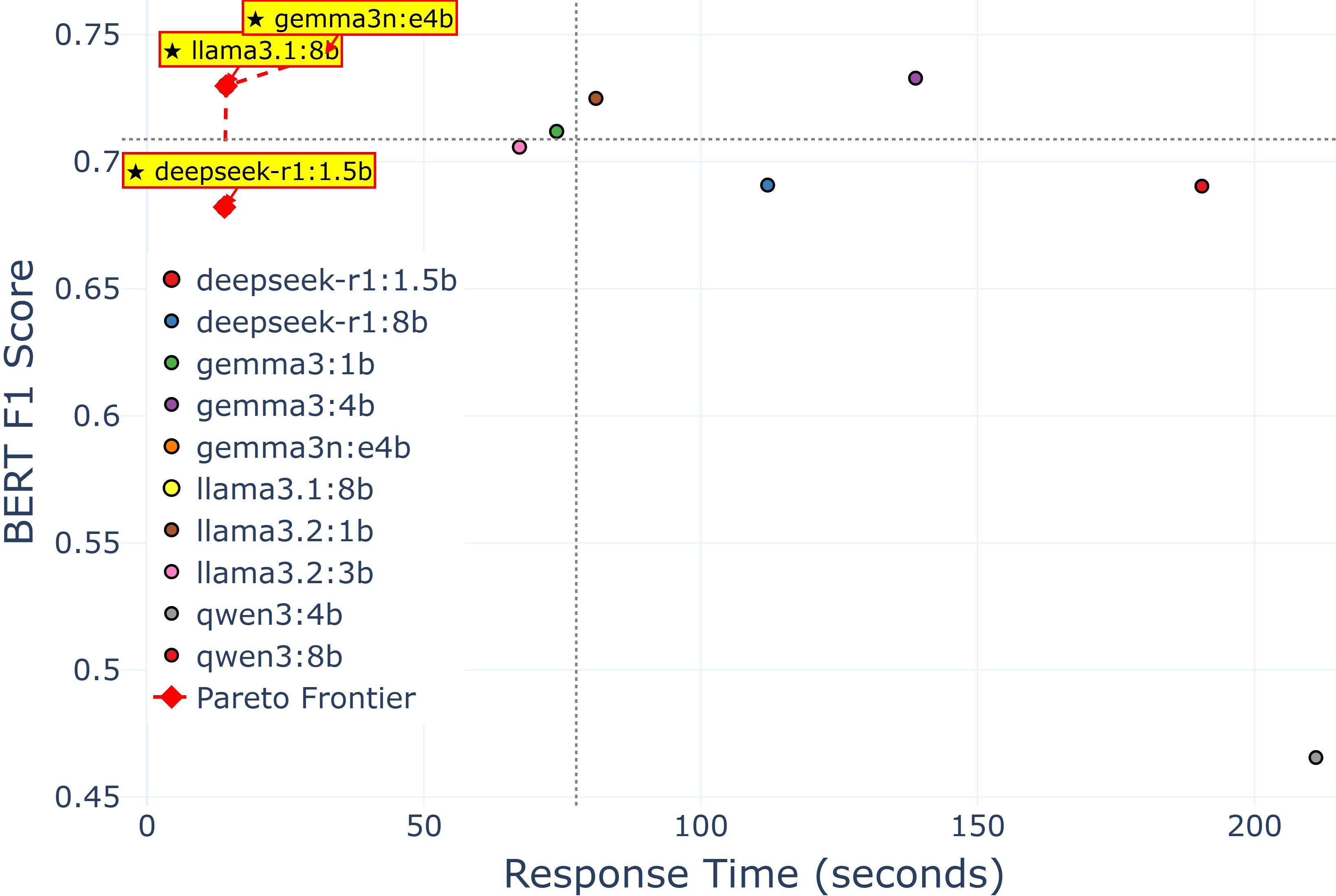}
    \caption{Speed Quality Efficiency}
    \label{fig:eval_speed_quality}
  \end{minipage}
\end{figure}

Technical evaluation employed BERTScore precision, recall, and F1 to measure semantic similarity with reference answers.
As shown in Fig.~\ref{fig:eval_precision_recall}, the upper-right cluster contains the most reliable small- and medium-sized models, with precision between $0.60$–$0.75$ and recall between $0.70$–$0.76$. \texttt{gemma3n:e4b} achieved the highest precision ($0.75$) with recall $0.74$.
\texttt{Gemma3:4b} and \texttt{llama3.1:8b} exhibited similar performance, while \texttt{llama3.1:1b} and \texttt{llama3.2:3b} favored recall over precision.
\texttt{Deepseek-r1:1.5b}, \texttt{deepseek-r1:8b}, and \texttt{qwen3:8b} performed moderately (precision $0.62$–$0.64$, recall $\approx 0.75$).
\texttt{Qwen3:4b} was a clear outlier with substantially lower scores (precision $0.43$, recall $0.51$).

Fig.~\ref{fig:eval_speed_quality} relates BERT F1 to query response time. Dashed guides indicate a latency threshold ($\sim 80$s) and a quality bar (F1 $\sim 0.71$). The target zone lies in the upper-left quadrant. \texttt{gemma3n:e4b} and \texttt{llama3.1:8b} were Pareto-optimal~\cite{b29}, achieving strong accuracy (F1 $0.74$–$0.75$) with moderate latency ($\sim 30$s). \texttt{deepseek-r1:1.5b} was the fastest ($15$–$20$s) but less accurate (F1 $0.69$–$0.70$). \texttt{gemma3:4b} reached F1 $\sim 0.73$ at much higher latency ($\sim 145$s). In contrast, \texttt{qwen3:8b} and \texttt{deepseek-r1:8b} fell below both thresholds, while \texttt{qwen3:4b} again underperformed. Overall, \texttt{gemma3n:e4b} and \texttt{llama3.1:8b} provided the best balance, with \texttt{deepseek-r1:1.5b} offering a low-latency alternative when moderate accuracy is acceptable.

\begin{figure}[!t]
  \centering
  \begin{subfigure}[t]{0.48\columnwidth}
    \centering
    \includegraphics[width=\linewidth]{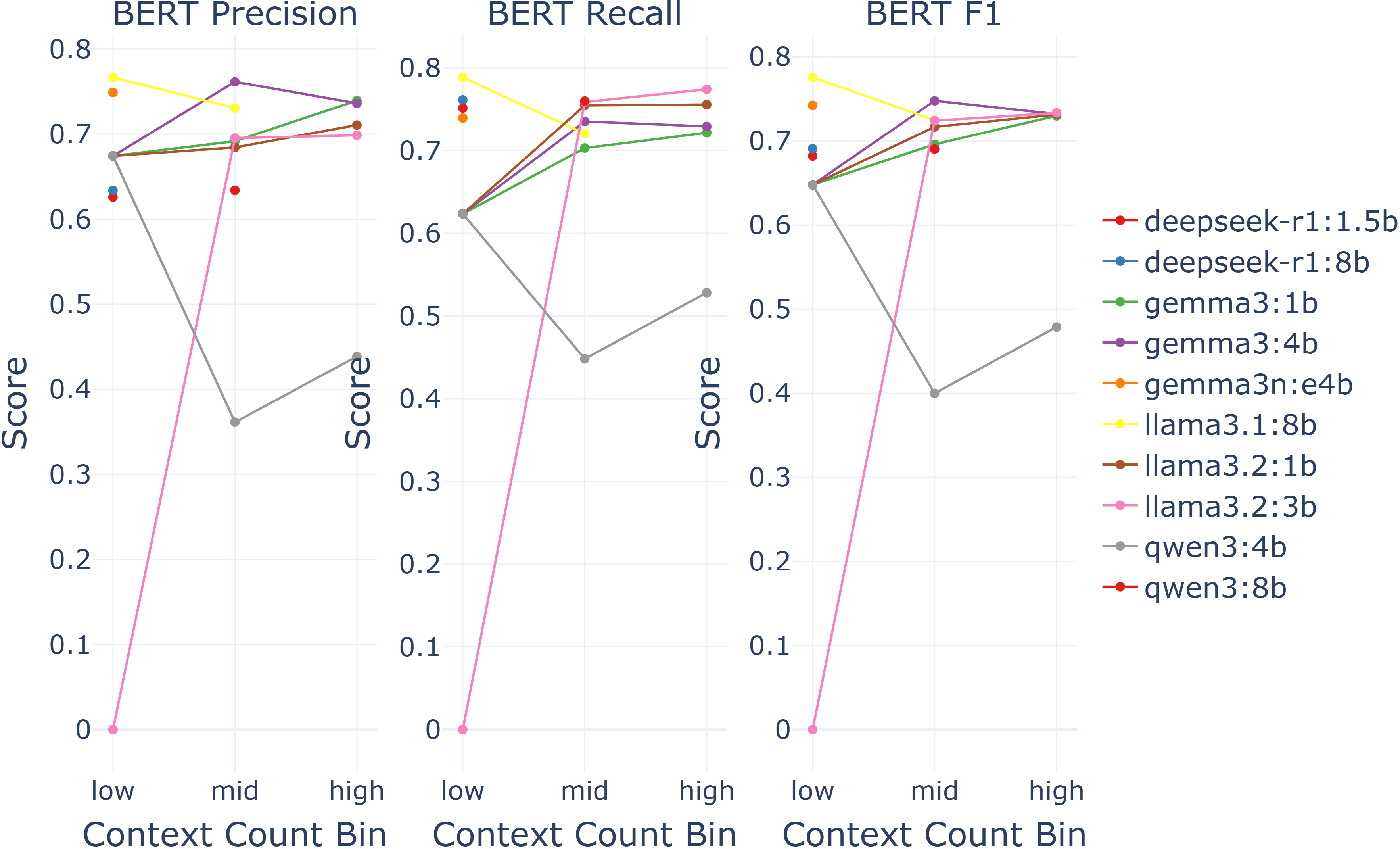}
    \caption{Reading Progress Robustness}
    \label{fig:eval_reading_progress_robustness}
  \end{subfigure}\hfill
  \begin{subfigure}[t]{0.48\columnwidth}
    \centering
    \includegraphics[width=\linewidth]{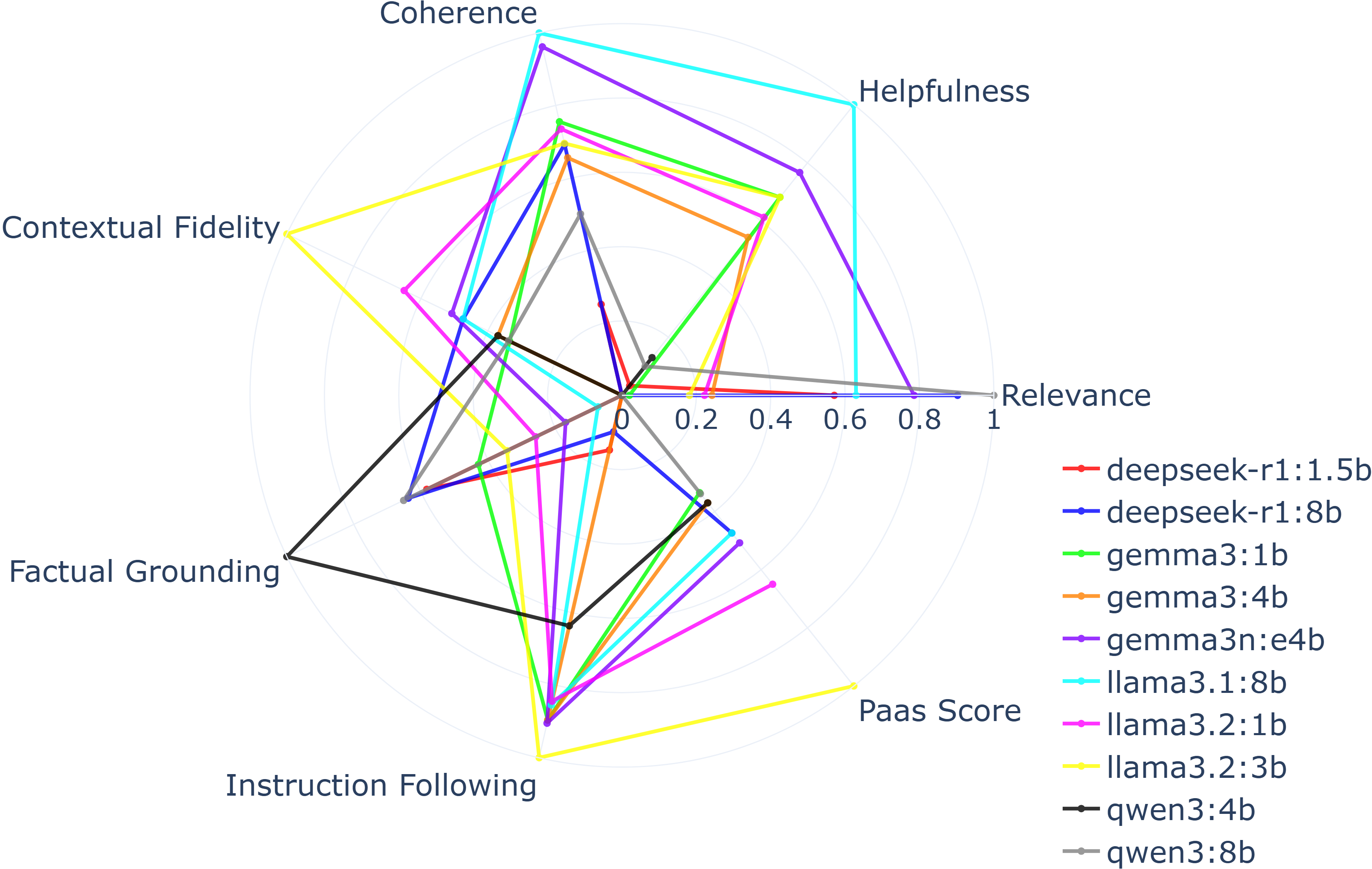}
    \caption{Qualitative Metric}
    \label{fig:eval_qualitative_metric}
  \end{subfigure}
  \caption{Robustness and Qualitative Radar}
  \label{fig:eval_bundle}
\end{figure}

Fig.~\ref{fig:eval_reading_progress_robustness} shows performance across progress bins: Low (1–2 passages), Mid (3–4), and High (5–6). Missing markers indicate that no items fell into a given bin due to either retrieval limitations or runner/API errors. \texttt{gemma3n:e4b}, \texttt{gemma3:4b}, and \texttt{llama3.1:8b} remained stable across bins, with precision $0.73$–$0.77$, recall $0.71$–$0.76$, and F1 $0.73$–$0.75$. These results suggest that even with minimal evidence, the models produced reliable answers, and additional passages did not degrade quality. \texttt{deepseek-r1:1.5b} and \texttt{deepseek-r1:8b} tracked slightly below but consistently across bins. By contrast, \texttt{qwen3:4b} showed a sharp decline at Mid (F1 $\sim 0.40$), recovering only partially at High (F1 $0.47$–$0.52$). \texttt{llama3.2:3b} produced no Low bin results).
But, when data was available, its F1 aligned with the $\sim 0.74$ band.
Even with limited evidence (1–2 passages), models maintained stable F1, reflecting adherence to the instruction not to speculate.

For qualitative evaluation, we employed an LLM-as-judge framework (Fig.~\ref{fig:eval_qualitative_metric}) across five dimensions: contextual fidelity, relevance, helpfulness, coherence, and instruction following. This approach provided a human-like assessment of response quality while ensuring evaluation consistency. The radar plots reveal that \texttt{gemma3n:e4b} and \texttt{llama3.1:8b} achieve consistently high scores across most dimensions, especially coherence and contextual fidelity, aligning with their strong precision–recall performance. \texttt{deepseek-r1:1.5b} delivers high instruction following but lags in coherence and helpfulness, reflecting its lower BERT F1 despite faster responses. In contrast, \texttt{qwen3:4b} underperforms across all axes, particularly in factual grounding and coherence, consistent with its outlier status in quantitative metrics.
These findings inform that models with balanced technical accuracy also tend to produce more user-satisfactory outputs in qualitative dimensions.

Fig.~\ref{fig:eval_qualitative_metric} summarizes the judgments averaged over the benchmark dimensions.
\texttt{Llama3.1:8b} and \texttt{gemma3n:e4b} form the best overall shapes with high helpfulness, coherence, and instruction following and competing contextual fidelity.
\texttt{Llama3.2:3b} has a high instruction following and contextual fidelity (strong adherence to prompts and book grounding), but is weaker on relevance (consistent with occasional over-literal responses).
\texttt{Deepseek-r1:8b} scores high on relevance and factual grounding, but trials on helpfulness.
\texttt{Qwen3:4b} shows high factual grounding but an overall smaller polygon all around, matching its error profile and low precisions seen in the earlier diagrams.
Overall, these readings align with those of the quantitative metrics.
Models that are coherent and helpful also sit near the precision-recall frontier, while models with generation errors have visibly collapsed polygons.

%% file: chapters/06_Conclusion.tex
The paper introduced MeReader, an offline, privacy-preserving, and progress-aware reading companion that constrains both retrieval and generation to a reader's current position in the narrative.
Rather than answering "about the book," it recalls and clarifies without revealing future content or substituting for the act of reading.
Empirically, across three novels and 600 items (300 comprehension, 300 spoiler checks), we observe a consistent quality-latency frontier (Figure~\ref{fig:eval_speed_quality}); \texttt{gemma3n:e4b} and \texttt{llama3.1:8b} achieve the best balance (Bert F1 $\approx 0.74$ -$0.75$ at $\approx 30$s/item, while \texttt{deepseek-r1:1.5b} offers a low-latency alternative ($\approx 15$-$20$s at slightly lower quality (F1 $\approx 0.69$-$0.70$).
Precision-recall distributions (Figure~\ref{fig:eval_precision_recall}) place \texttt{gemma3n:e4b} at the upper frontier (P$\approx0.75$, R$\approx0.74$) with \texttt{llama3.1:8b} and \texttt{gemma3:4b} close behind.
Robustness vs. context availability (Figure~\ref{fig:eval_reading_progress_robustness}) shows stable performance even with $1$-$2$ passages, indicating the system's retrieval pipeline uses evidence that is both compact and sufficient.
Qualitative judging (Figure~\ref{fig:eval_qualitative_metric}) aligns with its quantitative counterparts; the Pareto-efficient models also lead on coherence, contextual fidelity, and helpfulness, i.e., they \textit{feel} better to use, not just score better.

The central novelty is to treat spoiler prevention as a first-class citizen, not an after-the-fact filter.
MeReader's progress gates ensure that generation cannot cite beyond what the reader has already read.
Taken together, the results provide a blueprint for a local, trustworthy reading support on commodity hardware by (1) keeping data and inference on device, (2) incorporating progress metadata into both retrieval and prompting, and (3) targeting frontier-efficient models that deliver strong comprehension at a tolerable latency.
For readers, this design encourages active engagement (asking, checking, connecting) instead of passive consumption.

Our study spans only three (classic) English novels, ten small/medium models when it comes to consumer hardware, and a single workstation configuration.
These may not be enough to form a comprehensive conclusion on said models and the system we used, but are enough to recognise patterns and the possible conclusions we can derive.
The metrics employ BERTScore and an LLM-as-judge, which both carry known biases and do not fully capture long-range discourse reasoning.
The reference answers used, by \texttt{Gemini 2.5 Pro}, were short and may under-reward creative, but at the same time faithful, paraphrasing.
Lastly, the spoiler protocol captures explicit future content but not all forms of implicit foreshadowing.

For future work, we see four directions to improve upon. First, scale evaluation by testing more models on a wider set of prompts/questions and a larger, more diverse pool of books spanning languages and genres; a broader dataset will enable stronger conclusions about MeReader’s effectiveness. Second, broaden generality and accessibility by extending support to multilingual texts and non-fiction. Third, conduct user studies to compare MeReader against baselines, measuring learning outcomes, enjoyment, and engagement. Finally, strengthen evaluation methodology by replacing the judge-as-LLM with human audits.

MeReader shows that progress-aware RAG can deliver spoiler-safe, high-quality reading guidance without leaving the device.
By aligning retrieval, prompting, and evaluation with the reader's position, the system augments, rather than replaces, the act of reading.
We view this as a practical step towards local AI companions for literary engagement and an invitation to the community to refine the blueprint along whatever axes they see fit.

For reviewing the source code of this project and the data used for analysis, please consult the GitHub repository \href{https://github.com/WhiteHades/mereader}{\path{github.com/WhiteHades/mereader}}~\cite{b30}.

%% file: references.bib
@misc{b1,
  author = {Naomi S. Baron},
  title = {Reading in a digital age},
  howpublished = {Kappan Online},
  note = {Accessed: May 29, 2025. [Online]. Available: \url{https://kappanonline.org/reading-digital-age/}}
}

@misc{b2,
  title = {{ChatGPT}},
  author = {{OpenAI}},
  note = {Accessed: May 29, 2025. [Online]. Available: \url{https://openai.com/chatgpt/overview/}}
}

@misc{b3,
  title = {{Llama}},
  author = {{Meta}},
  note = {Accessed: May 29, 2025. [Online]. Available: \url{https://www.llama.com/}}
}

@article{b6,
  author = {C. Chen and A. Leitch},
  title = {LLMs as Academic Reading Companions: Extending HCI Through Synthetic Personae},
  journal = {arXiv preprint arXiv:2403.19506},
  year = {2024},
  doi = {10.48550/arXiv.2403.19506}
}

@article{b7,
  author = {W. Kryściński and N. Rajani and D. Agarwal and C. Xiong and D. Radev},
  title = {BookSum: A Collection of Datasets for Long-form Narrative Summarization},
  journal = {arXiv preprint arXiv:2105.08209},
  year = {2022},
  doi = {10.48550/arXiv.2105.08209}
}

@article{b8,
  author = {S. E. Spatharioti and D. M. Rothschild and D. G. Goldstein and J. M. Hofman},
  title = {Comparing Traditional and LLM-based Search for Consumer Choice: A Randomized Experiment},
  journal = {arXiv preprint arXiv:2307.03744},
  year = {2023},
  doi = {10.48550/arXiv.2307.03744}
}

@article{b9,
  author = {V. Karpukhin and others},
  title = {Dense Passage Retrieval for Open-Domain Question Answering},
  journal = {arXiv preprint arXiv:2004.04906},
  year = {2020},
  doi = {10.48550/arXiv.2004.04906}
}

@article{b10,
  author = {B. Reichman and L. Heck},
  title = {Dense Passage Retrieval: Is it Retrieving?},
  journal = {arXiv preprint arXiv:2402.11035},
  year = {2024},
  doi = {10.48550/arXiv.2402.11035}
}

@article{b11,
  author = {Y. Qu and others},
  title = {RocketQA: An Optimized Training Approach to Dense Passage Retrieval for Open-Domain Question Answering},
  journal = {arXiv preprint arXiv:2010.08191},
  year = {2021},
  doi = {10.48550/arXiv.2010.08191}
}

@misc{b12,
  title = {{Nomic Embed Text V2: An Open Source, Multilingual, Mixture-of-Experts Embedding Model}},
  author = {{Nomic AI}},
  note = {Accessed: Jun. 04, 2025. [Online]. Available: \url{https://www.nomic.ai/blog/posts/nomic-embed-text-v2}}
}

@article{b13,
  author = {X. Wu and C. Gao and Z. Lin and J. Han and Z. Wang and S. Hu},
  title = {InfoCSE: Information-aggregated Contrastive Learning of Sentence Embeddings},
  journal = {arXiv preprint arXiv:2210.06432},
  year = {2022},
  doi = {10.48550/arXiv.2210.06432}
}

@article{b14,
  author = {Y. Gao and others},
  title = {Retrieval-Augmented Generation for Large Language Models: A Survey},
  journal = {arXiv preprint arXiv:2312.10997},
  year = {2024},
  doi = {10.48550/arXiv.2312.10997}
}

@article{b15,
  author = {X. Wang and others},
  title = {Searching for Best Practices in Retrieval-Augmented Generation},
  journal = {arXiv preprint arXiv:2407.01219},
  year = {2024}
}

@inproceedings{b16,
  author = {W. Zhao and Z. Deng and S. Yadav and P. S. Yu},
  title = {Heterogeneous Knowledge Grounding for Medical Question Answering with Retrieval Augmented Large Language Model},
  booktitle = {Companion Proceedings of the ACM Web Conference 2024 (WWW '24)},
  pages = {1590--1594},
  year = {2024},
  publisher = {ACM},
  doi = {10.1145/3589335.3651941}
}

@article{b17,
  author = {L. Cabello and C. Martin-Turrero and U. Akujuobi and A. Søgaard and C. Bobed},
  title = {MEG: Medical Knowledge-Augmented Large Language Models for Question Answering},
  journal = {arXiv preprint arXiv:2411.03883},
  year = {2025},
  doi = {10.48550/arXiv.2411.03883}
}

@article{b18,
  author = {C. Mitra and M. Miroyan and R. Jain and V. Kumud and G. Ranade and N. Norouzi},
  title = {RetLLM-E: Retrieval-Prompt Strategy for Question-Answering on Student Discussion Forums},
  journal = {Proceedings of the AAAI Conference on Artificial Intelligence},
  volume = {38},
  number = {21},
  year = {2024},
  doi = {10.1609/aaai.v38i21.30368}
}

@article{b19,
  author = {Z. Levonian and others},
  title = {Retrieval-augmented Generation to Improve Math Question-Answering: Trade-offs Between Groundedness and Human Preference},
  journal = {arXiv preprint arXiv:2310.03184},
  year = {2023},
  doi = {10.48550/arXiv.2310.03184}
}

@article{b20,
  author = {Z. Jiang and others},
  title = {Active Retrieval Augmented Generation},
  journal = {arXiv preprint arXiv:2305.06983},
  year = {2023},
  doi = {10.48550/arXiv.2305.06983}
}

@article{b21,
  author = {J. Xu and others},
  title = {On-Device Language Models: A Comprehensive Review},
  journal = {arXiv preprint arXiv:2409.00088},
  year = {2024},
  doi = {10.48550/arXiv.2409.00088}
}

@misc{b22,
  title = {{SQLite Home Page}},
  author = {{SQLite}},
  note = {Accessed: Jun. 04, 2025. [Online]. Available: \url{https://www.sqlite.org/}}
}

@misc{b23,
  title = {{Qdrant - Vector Database}},
  author = {{Qdrant}},
  note = {Accessed: Jun. 04, 2025. [Online]. Available: \url{https://qdrant.tech/}}
}

@misc{b24,
  title = {{Ollama}},
  author = {{Ollama}},
  note = {Accessed: Jun. 04, 2025. [Online]. Available: \url{https://ollama.com}}
}

@inproceedings{b25,
  author = {F. Mireshghallah and A. Uniyal and T. Wang and D. Evans and T. Berg-Kirkpatrick},
  title = {An Empirical Analysis of Memorization in Fine-tuned Autoregressive Language Models},
  booktitle = {Proceedings of the 2022 Conference on Empirical Methods in Natural Language Processing (EMNLP)},
  pages = {1816--1826},
  year = {2022},
  doi = {10.18653/v1/2022.emnlp-main.119}
}

@article{b26,
  author = {D. Ippolito and others},
  title = {Preventing Verbatim Memorization in Language Models Gives a False Sense of Privacy},
  journal = {arXiv preprint arXiv:2210.17546},
  year = {2023},
  doi = {10.48550/arXiv.2210.17546}
}

@inproceedings{b27,
  author = {H. Zhou and S. Boovaraghavan and M. Goel and Y. Agarwal},
  title = {On-Device Speech Filtering for Privacy-Preserving Acoustic Activity Recognition},
  booktitle = {Proceedings of the 30th Annual International Conference on Mobile Computing and Networking (ACM MobiCom '24)},
  pages = {1802--1804},
  year = {2024},
  doi = {10.1145/3636534.3698865}
}

@misc{b28,
  title = {What is BM25 (Best Matching 25) Algorithm?},
  author = {{GeeksforGeeks}},
  note = {Accessed: Jun. 11, 2025. [Online]. Available: \url{https://www.geeksforgeeks.org/what-is-bm25-best-matching-25-algorithm/}}
}

@misc{b29,
  title = {Pareto Frontier - an overview},
  author = {{ScienceDirect}},
  note = {Accessed: Sep. 10, 2025. [Online]. Available: \url{https://www.sciencedirect.com/topics/engineering/pareto-frontier}}
}

@misc{b30,
  author = {Mohammed Efaz},
  title = {{MeReader: Progress-Aware, On-Device Reading Companion}},
  howpublished = {GitHub repository},
  year = {2025},
  note = {\url{https://github.com/WhiteHades/mereader}}
}

@misc{b31,
  title = {Nineteen Eighty-Four},
  author = {{George Orwell}},
  note = {Accessed: Nov. 04, 2025. [Online]. Available: \url{https://en.wikipedia.org/wiki/Nineteen_Eighty-Four}}
}

@misc{b32,
  title = {Of Mice and Men},
  author = {{John Steinbeck}},
  note = {Accessed: Nov. 04, 2025. [Online]. Available: \url{https://en.wikipedia.org/wiki/Of_Mice_and_Men}}
}

@misc{b33,
  title = {The Death of Ivan Ilyich},
  author = {{Leo Tolstoy}},
  note = {Accessed: Nov. 04, 2025. [Online]. Available: \url{https://en.wikipedia.org/wiki/The_Death_of_Ivan_Ilyich}}
}

@article{verge_kindle,
  title = {Kindle app now answers questions about the book you’re reading},
  author = {Preston, Dominic},
  journal = {The Verge},
  year = {2025},
  month = {Dec},
  url = {https://www.theverge.com/news/844538/kindle-app-ask-this-book-ai-ios}
}

@article{engadget_kindle,
  title = {Kindle's in-book {AI} assistant can answer all your questions without spoilers},
  author = {Malone, Kirra},
  journal = {Engadget},
  year = {2025},
  url = {https://www.engadget.com/ai/kindles-in-book-ai-assistant-can-answer-all-your-questions-without-spoilers-190609961.html}
}

@article{verge_pdfs,
  title = {Adobe {Acrobat} gets a generative {AI} assistant to chat with your documents},
  author = {Weatherbed, Jess},
  journal = {The Verge},
  year = {2024},
  month = {Feb},
  url = {https://www.theverge.com/2024/2/20/24077217/adobe-acrobat-generative-ai-assistant-chatbot-pdf-document}
}

@misc{github_koreader_plugin,
  title = {Assistant: {AI} Helper Plugin for {KOReader}},
  author = {Omer Faruq},
  year = {2026},
  howpublished = {\url{https://github.com/omer-faruq/assistant.koplugin}},
  note = {GitHub repository}
}

@misc{github_koreader,
  title        = {{KOReader}: An ebook reader application supporting PDF, DjVu, EPUB, FB2 and many more formats},
  author       = {{KOReader Contributors}},
  year         = {2026},
  howpublished = {\url{https://github.com/koreader/koreader}},
  note         = {GitHub repository}
}
